\documentclass{epl}

\title{The vibrational dynamics of vitreous silica: 
Classical force fields vs. first-principles}
\shorttitle{}
\author{Magali Benoit and Walter Kob}
\institute{ Laboratoire des Verres, Universit\'e Montpellier II, 
  34090 Montpellier, France 
}

\begin{document}

\maketitle

\vspace*{-10mm}

\begin{abstract}
We compare the vibrational properties of model SiO$_2$ glasses generated
by molecular-dynamics simulations using the effective force field of van
Beest {\it et al.} (BKS) with those obtained when the BKS structure is relaxed
using an {\it ab initio} calculation in the framework of the density
functional theory. We find that this relaxation significantly improves
the agreement of the density of states with the experimental result. For
frequencies between 14 and 26 THz the nature of the vibrational modes
as determined from the BKS model is very different from the one from the
{\it ab initio} calculation, showing that the interpretation of the
vibrational spectra in terms of calculations using effective potentials can be very misleading.

\end{abstract}

\vspace*{-10mm}
\section{Motivation}
Understanding the microscopic properties of vibrational excitations
in disordered systems is a long standing challenge in basic physics as well
as in material science since the lack of positional order makes both the
experimental study and the theoretical interpretations of the
results very difficult. For instance the mechanism leading to the existence the so-called boson peak present in many glasses is the subject
of a long standing debate, and the reason for the presence of the
D$_1$ and D$_2$ lines in the Raman spectra of amorphous SiO$_2$ has remained 
unclear for a long time
~\cite{winterling75,galeener76,buchenau86,foret96,benassi96,
wischnewski98,hehlen00,pilla00}.

In principle molecular dynamics (MD) computer simulations 
overcome these difficulties since one has direct access to all
the necessary microscopic information. Therefore in recent
years many studies of this kind have been carried out with the aim of
shedding some light on the nature of these vibrational excitations,
in particular for the case of silica, the paradigm of network
forming glasses~\cite{pasquarello_prl95,wilson96,guillot97,
pasquarello_science97,taraskin97,elliott_197,elliott_297,pasquarello_Sqw98,
pasquarello_raman98,uchino_raman00,horbach01}. Due to the large
computational costs of such simulations the vast majority of them were
done with {\it effective} classical force fields, i.e. potentials which
were optimized to reproduce certain (somewhat arbitrarily chosen)
experimental features of SiO$_2$. It is clear that the reliability
of the results of these investigations depends crucially on whether
or not the interactions used are sufficiently accurate to allow
a faithful description of the real material. Hence a considerable
effort has been made to check that the models used  do  reproduce
the salient structural and dynamical features of real silica. In
these studies it has been shown that the classical force
fields employed are indeed able to give a good description of quantities
like the structure factor, the diffusion constant or viscosity,
etc.~\cite{garofalini82,vashishta93,dellavalle94,vollmayr96,guillot97,
horbach99}, so their use in investigations of the vibrational
properties also appears a reasonable undertaking. Nevertheless, it was found that
certain features of the vibrational density of states (DOS) are not
well reproduced by these models and more sophisticated calculations,
such as the numerical studies based on first-principles, seem to
be required~\cite{zotov99}, in agreement with conclusions drawn
for the case of liquids~\cite{silvestrelli97}. Pasquarello {\it et
al.} showed that using an {\it ab initio} scheme it is possible to
reproduce many structural, electronic and vibrational properties of real
silica~\cite{pasquarello_prl95,pasquarello_science97,pasquarello_Sqw98,pasquare
llo_raman98}.
However this approach suffers from its heavy computational cost which
restricts this type of calculation to the study of very small systems
with a rather low statistical accuracy.

One possibility to overcome this limitation, at least partially, is to
use a combined approach which consists in generating a glass using an
effective potential and subsequently refining the structure obtained by
means of first-principles \cite{EPJB00}. In previous work we showed that
the {\it structure} of vitreous SiO$_2$ generated using the effective
force field by van Beest, Kramer and van Santen (BKS)~\cite{BKS90} is
only modified weakly by a first-principles calculation, thus validating
the structural model generated with this potential.

In contrast to this we show in this letter that the DOS of a SiO$_2$ glass
generated by classical MD simulations using the BKS potential is strongly
modified by using an {\it ab initio} treatment of the forces, and that
this treatment leads to a much better agreement with experimental results. 
Moreover,
in a large frequency range, the nature of the excitations as determined
from the effective potential differs significantly from the one determined
from the {\it ab initio} forces thus raising doubts as to the detailed
analysis of the nature of the vibrational excitations determined from
the BKS force field.

\section{Simulation details}
Molecular-dynamics simulations were done using the BKS potential
on systems containing 26 SiO$_2$ units at the experimental density
(2.2 g/cm$^{3}$). For this we used the velocity form of the Verlet
algorithm with a time step of 1.63 fs. Three different samples were
generated by quenching liquids well-equilibrated at 3500 K to 300 K,
using three different cooling rates: $5\cdot 10^{12}$ K/s, $3\cdot
10^{11}$ K/s, and $7\cdot 10^{10}$ K/s. The  glasses obtained this way (which are 
non-equilibrium structures) were annealing for 70 ps at 300 K, and
subsequently quenched to 0 K, at which their dynamical matrices were
evaluated and diagonalized in order to obtain the vibrational frequencies
and the corresponding (normalized) eigenmodes. In parallel the final
atomic coordinates and velocities after the annealing at 300 K were used
as initial coordinates and velocities for short  ($\approx$ 0.12 ps)
{\it ab initio} molecular-dynamics simulations of the Car-Parrinello
type \cite{CP85}, using the CPMD code \cite{CPMD95}. The technical
details of these simulations were identical to the ones described
in Ref. \cite{EPJB00}. At the end of these simulations the structures
of the three glasses were relaxed to 0 K and the dynamical matrices
were computed by evaluating the second derivatives of the total energy
with respect to atomic displacements by taking finite differences of
the atomic forces. Subsequently the vibrational frequencies and the
corresponding eigenmodes were obtained from these matrices. Hence we
obtained $g(\omega)$, the {\em true} DOS for this system. Note that
although the cooling rates are high and the system size is small, the
DOS depends only weakly on these parameters~\cite{vollmayr96}.

In the following, the quantities computed by means of classical
molecular-dynamics simulations using the BKS potential and the CPMD code
will be labeled ``BKS" and ``CP", respectively.

\section{Results}
\begin{figure}
\onefigure[width=8.5cm]{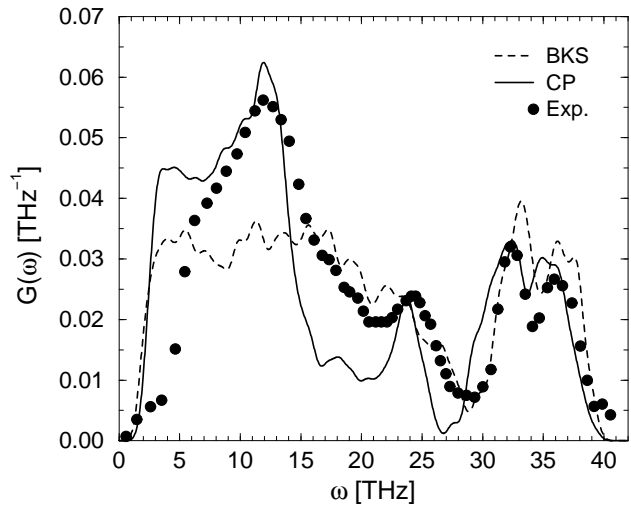}
\caption{Frequency dependence of the effective density of states
$G(\omega)$ calculated from {\it ab initio} (solid line) and classical
(dashed line) molecular dynamics simulations and compared to neutron
scattering experiments from Ref.~\cite{carpenter85} (filled circles). The
calculated $G(\omega)$ have been obtained using the experimental values
for the neutron scattering length factors $b_{\rm Si} = 4.149$ fm and
$b_{\rm O} = 5.803$ fm and a Gaussian broadening of width $2\sigma=1.05$ THz.}
\label{fig1}
\end{figure}

Since we found that the DOS from the three different cooling rates
are identical to within the statistical error - which is relatively large
due to the small system size -, we decided to treat the three glasses
as independent statistical samples and analyzed the three sets of
vibrational frequencies/modes together. The resulting vibrational
DOS was used to compute an {\em effective} neutron scattering
cross section $G(\omega)=C(\omega)g(\omega)$. This was done by
using the incoherent approximation and by calculating $C(\omega)$
as in Ref.~\cite{elliott_197}.  We note that the correction functions
$C(\omega)$ for the BKS and CP are very similar and hence differences in
the respective $G(\omega)$ are mainly due to differences in the respective
$g(\omega)$. In Fig.~\ref{fig1} we compare the $G(\omega)$ obtained
and we see that at intermediate frequencies the two curves are very
different. In particular we see that the CP curve has a pronounced peak
at around 12 THz and a smaller one at around 24 THz. Finally there is
a small peak at 18 THz, the so-called D$_2$ line, which is due to a
ring of size three. Overall the CP curve is in very good agreement with
previous investigations~\cite{pasquarello_raman98}.  All these features
are missing in the BKS curve, despite the good agreement between CPMD and
BKS with regard to structural properties.  Also included is the result of
a neutron scattering experiment by Carpenter and Price ~\cite{carpenter85}
and from the reasonable agreement between this curve and the one from
the CP calculation we conclude that the latter is reliable.  [ Note
that i) there is no fit parameter whatsoever, and ii) the lack of a
small peak at around 4 THz in the experimental data is  related to the
insufficient experimental resolution~\cite{wischnewski98} ].  Hence we
conclude from this figure that the DOS as calculated from the BKS model
is not very reliable at intermediate frequencies, in agreement with
Ref.~\cite{guillot97}. Note that similar discrepancies between experiments
and simulations with various effective interactions have already been
observed in previous studies~\cite{vashishta93,vollmayr96,elliott_197,
elliott_297,guillot97} and hence we conclude that  many other types
of force fields also lead to a density of states which is not trustworthy
and that most probably the conclusions drawn in this paper hold 
for these other potentials as well. However, the good agreement between the
CP and the experimental DOS clearly shows that a simple refinement of
the BKS model glass by a first-principles calculation is sufficient to
significantly improve the vibrational density of states.

\begin{figure}
\onefigure[width=9.0cm]{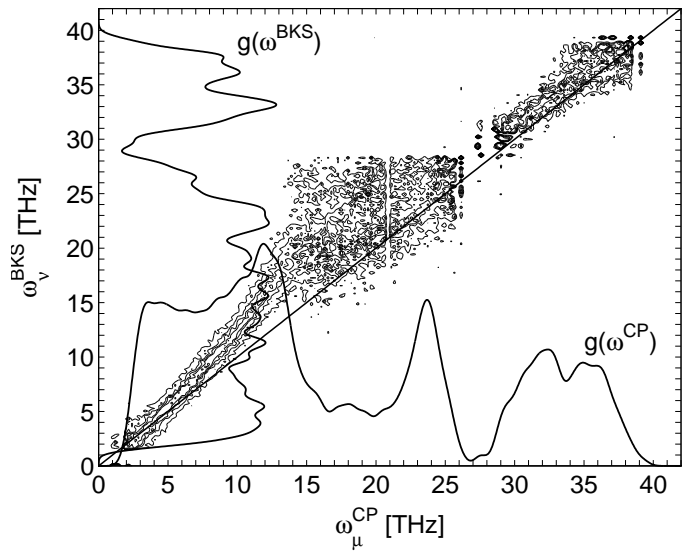} 
\caption{Contour plot of $H(\omega^{\rm BKS}_{\nu}, \omega^{\rm CP}_{\mu})$, 
the
projection of the BKS eigenmodes onto the CP modes (see Eq. 
(\protect\ref{eq1})). The
solid lines are the DOS $g(\omega)$ as determined from the BKS and the CP 
glasses respectively,
scaled in order to fit into the graph. [Contour levels: 0.06, 0.1, 0.15, 0.2, 
0.3, 0.5, 0.8 1.0] }
\label{fig2}
\end{figure}

Having shown that the effective DOS from the BKS differs significantly
from the one from the CP we now investigate this difference in more
detail by comparing the corresponding eigenmodes. Since the number
of eigenmodes for the BKS and CP cases is the same, Fig.~\ref{fig1}
implies that there is a considerable reshuffling of the modes from one
frequency to another if the force field is switched. In order to check
whether at a given frequency $\omega_{\nu}^{\rm BKS}$ ($\nu =1,\ldots,3 \times 
3N$,
and $N$ is the number of atoms in each of the three samples) 
the nature of a given mode ${\bf e}^{\rm BKS}(\omega_{\nu}^{\rm BKS})$ 
from the BKS system is similar to one of the
CP modes we calculated the projections of the latter onto the former modes:
\begin{equation}
H(\omega^{\rm BKS}_{\nu}, \omega^{\rm CP}_{\mu}) =  
\left| {\bf e}^{\rm BKS}(\omega^{\rm BKS}_{\nu}) \cdot 
{\bf e}^{\rm CP}(\omega^{\rm CP}_{\mu})  \right|
\quad {\rm with} \quad \nu, \mu=1,\ldots,3 \times 3N .
\label{eq1}
\end{equation}
Here ${\bf e}^{\rm CP}(\omega^{\rm CP}_{\mu})$ is the eigenvector for the
CP system at frequency $\omega^{\rm CP}_{\mu}$.  With this definition,
$H$ varies between 0 and 1, since $|{\bf e}^{\rm BKS}(\omega^{\rm
BKS}_{\nu})| = |{\bf e}^{\rm CP}(\omega^{\rm CP}_{\mu})| = 1$. If the
nature of the modes in the BKS system were identical to the ones
of the CP system, $H$ would be a line of $\delta-$functions along the 
diagonal in the $(\omega_{\nu}^{\rm BKS},\omega^{\rm CP}_{\mu})$ plane. The 
frequency
dependence of $H(\omega^{\rm BKS}_{\nu}, \omega^{\rm CP}_{\mu})$ is
shown in Fig.~\ref{fig2} as a contour plot.

This plot shows a well-defined ridge close to the diagonal of the graph up
to $\approx$ 7 THz and between $\approx$ 28 and 39 THz. This means that the
overlap between the BKS and the CP vibrational modes is significant in
these frequency ranges, i.e. that at a given frequency the nature of the
mode is very similar. For 7 THz $\leq \omega^{\rm CP} \leq $ 13 THz we still
find a well defined ridge, but its location is above the diagonal. Hence
we see that although the BKS modes can still be  well represented by the CP
ones, they are shifted to slightly higher frequencies. Since the interval
7 THz $\leq \omega^{\rm BKS} \leq $ 21 THz is compressed into the range
7 THz $\leq \omega^{\rm CP} \leq $ 14 THz this leads to a significant larger
$g^{\rm CP}(\omega)$ in this range, in agreement with Fig.~\ref{fig1}.

In the range 16 THz $\leq \omega^{\rm BKS} \leq$ 28 THz we find that the BKS
modes have a significant overlap with CP modes that cover a large range
in $\omega^{\rm CP}$. This implies that in this frequency range the vibrational
dynamics of the BKS system is no longer realistic and that hence care
must be taken if one draws conclusions from the analysis of the modes
in this range.

In order to investigate whether in this frequency range the CP
eigenmodes can at least be written as a sum of only a {\it
few} BKS modes, we ordered for each frequency $\omega_{\mu}^{\rm CP}$
the overlaps $H(\omega_{\nu'}^{\rm BKS}, \omega_{\mu}^{\rm CP})$ in descending
order, i.e. $[H(\omega_1^{\rm BKS}, \omega_{\mu}^{\rm CP})]^2 \geq 
[H(\omega_2^{\rm BKS},\omega_{\mu}^{\rm CP})]^2
\geq \ldots$, and calculated the sum
\begin{equation}
S_J(\omega_{\mu}^{CP}) = 
\sum_{\nu'=1}^J \left[ H(\omega_{\nu'}^{\rm BKS}, \omega_{\mu}^{\rm CP}) 
\right]^2 \quad.
\label{eq2}
\end{equation}
\begin{figure}
\onefigure[width=13.5cm]{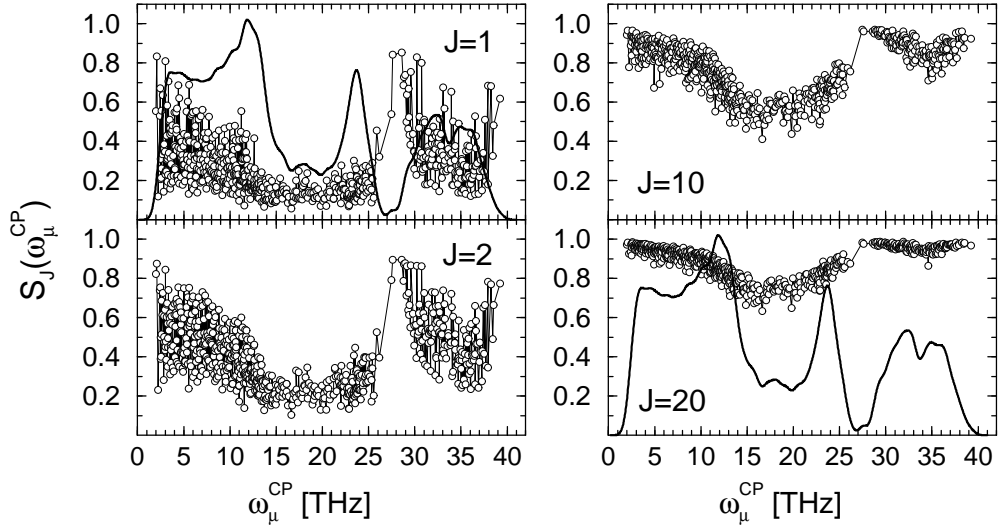} 
\caption{Circles: Sum of the $J$ largest overlaps between the BKS and CP
modes, as defined in Eq. (\protect\ref{eq2}), for each CP eigenfrequency
$ \omega_{\mu}^{\rm CP}$. Bold solid line: The CP-DOS (scaled in order to fit 
in
the graph).}
\label{fig3}
\end{figure}
\noindent
(Note that the sum of such eigenmodes is in general no longer an eigenmode.)
The resulting $S_J$ as a function of the CP vibrational frequencies
$\omega_{\mu}^{\rm CP}$ are presented in Fig.~\ref{fig3} for $J$=1, 2,
10, and 20.
This figure demonstrates that in general the CP modes are not well represented  
until the sum extends over at least 10 BKS eigenmodes. The only
exceptions are modes at very low frequency ($\leq$ 3 THz) and around
27, 29, and 38 THz. These latter modes show already a prominent peak
in $S_1(\omega_{\mu}^{\rm CP})$ and are at band edges and hence are
localized. Note that even for 3 THz $\leq \omega^{\rm CP} \leq$ 13 THz,
i.e. the frequency range where we find a strong correlation in
Fig.~\ref{fig2}, the BKS modes do not describe well the CP modes. In
the range 13 THz $\leq \omega^{\rm CP} \leq$ 26 THz one needs on the order
of 20 BKS modes in order to describe a CP mode, which shows that 
the former have not much in common with the latter.

\begin{figure}
\onefigure[width=12.5cm]{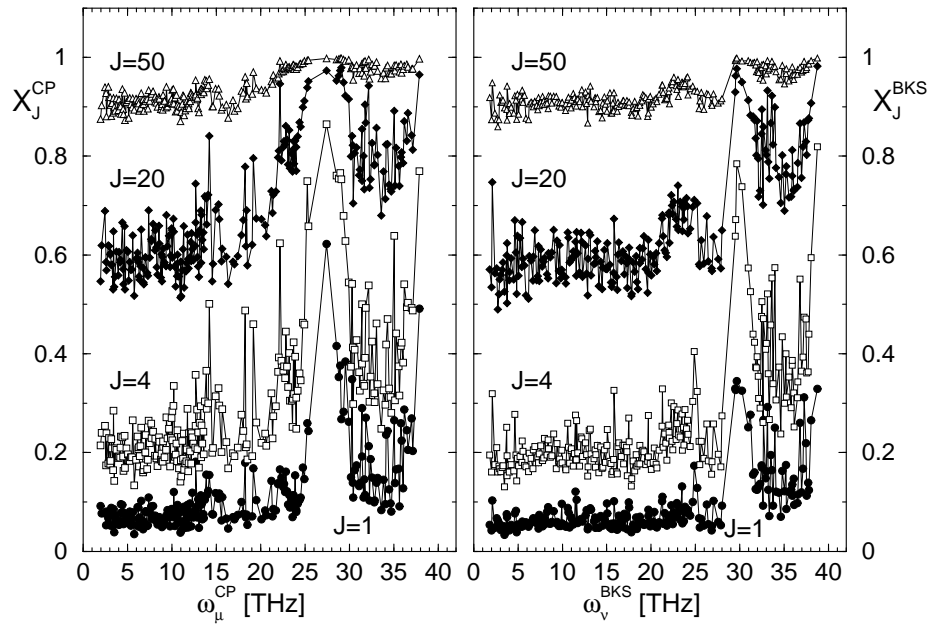} 
\caption{Sum of the $J$ largest components of an eigenvector with frequency
$\omega$ (see Eq. (\protect\ref{eq3})) for $J=$ 1, 4,  20 and 50.
Left panel: $X_J^{\rm CP}$; Right panel: $ X_J^{\rm BKS}$.}
\label{fig4}
\end{figure}

It is well known that the vibrational excitations of silica include
modes that are very localized as well as ones that are collective. In
order to see whether there is a difference in the ability of the BKS
potential to describe the one or the other type we have defined the
following quantities:
\begin{equation}
X^{\rm CP}_J(\omega^{\rm CP}_{\mu}) = 
\sum_{\alpha'=1}^J \sum_{i=1}^3 \left|e_{\alpha',i}^{\rm CP}(\omega^{\rm 
CP}_{\mu})\right|^2
\quad {\rm and} \quad
X^{\rm BKS}_J(\omega^{\rm BKS}_{\nu}) = 
\sum_{\alpha'=1}^J \sum_{i=1}^3 \left|e_{\alpha',i}^{\rm BKS}(\omega^{\rm 
BKS}_{\nu}) \right|^2
\label{eq3}
\end{equation}
where  $\alpha'$ is the atomic index, $i = x,y,z$  and  
the prime in the sum ($\alpha'$) indicates that we have made the ordering: 
$\sum_{i=1}^3  |e_{1,i}|^2 \geq \sum_{i=1}^3  |e_{2,i}|^2 \geq \cdots $. Thus 
$X^{\rm
CP}_J(\omega^{\rm CP}_{\mu})$ is the sum of the $J$ largest atomic 
displacements
for that eigenvector. The resulting $X^{\rm BKS}$ and $X^{\rm CP}$
are shown in Figure \protect\ref{fig4}, for values of $J$=1, 4, 20 and
50. (Note that for the sake of clarity the quantity is only shown for
one sample. The other two samples look qualitatively similar.)

Using this quantity, we can estimate the number of atoms participating in
a given vibrational mode. Firstly, we note that most of the modes with
frequencies below $\approx$ 22 THz have a ``collective" nature since
around 20 eigenvectors are needed to make 60 \% of $X$, independent of
the model used. Exceptions are CP modes at 14 and 18 THz which show a
pronounced peak with a height larger than 0.4 even for $J=4$. These
modes correspond to the D$_1$ and D$_2$ lines in the Raman spectrum
of vitreous SiO$_2$ that have been attributed to breathing modes of 3
and 4-membered rings \cite{pasquarello_raman98,uchino_raman00}. From
the curves it becomes clear that these modes are not present in the
BKS description, thus raising the problem of the interpretation of the
Raman spectrum of glassy SiO$_2$ using this model potential (see also
~\cite{zotov99}). For frequencies higher than $\approx$ 22 THz the nature
of the modes becomes more localized for the case of the CP model as well
as the BKS model. However, we note a clear difference in the $X_J^{\rm
BKS}$ and $X_J^{\rm CP}$ between 22 and 29 THz, where the BKS modes seem
to be more collective than the CP ones.  Finally we mention that one sees
a strong localization of the modes near the gap of the DOS and at around
39 THz, in agreement with the comment made in context of Fig.~\ref{fig3}.

\section{Conclusion}
We have shown that for the case of silica the DOS as predicted by an
effective force field can be improved considerably by relaxing the
configuration by means of an {\it ab initio} simulation, {\it without
a significant change in the structural properties}. For intermediate
frequencies the nature of the modes in the BKS system is very different
from the one in the CP system. In particular the effective force field is
not able to reproduce the nature of the modes attributed to the D$_1$
and D$_2$ lines in the Raman spectra. In view of the fact that the
BKS model was obtained by an {\it ab initio} calculation of a single
tetrahedron and a lattice dynamic simulation of a crystal to optimize
the elastic constants~\cite{BKS90}, it is not surprising that the model
does well at very high and very low frequencies, but is not reliable at
intermediate frequencies. Hence we conclude that despite the reliability
of effective potentials with regard to structural properties, it might
be that if one wants to investigate vibrational features in the whole
frequency range one would need an effective potential which is more accurate,
or a full {\it ab initio} calculation.

\acknowledgments

We thank J. Horbach, S. Ispas and R. Jullien for useful discussions.
Computations have been performed on the IBM SP3 of the national computer
center CINES in Montpellier, France.

\end{document}